\documentclass[prb,twocolumn,floatfix,amssymb,aps,showpacs,graphicx]{revtex4}
\usepackage{bm} 	
\usepackage{color}
\usepackage{amsmath}
\usepackage{epsfig}
\usepackage{amssymb}
\usepackage{amsfonts}

\begin{document}
\date{\today}
\title{\large\textbf{ Disordered  Quantum Spin Chains with Long-Range Antiferromagnetic Interactions}}
\author{N. Moure}
\email[]{mouregom@usc.edu}
\affiliation{Department of Physics and Astronomy  
University of Southern California,
 Los Angeles, CA 90089-0484}

\author{Hyun-Yong Lee}
\email[]{hyunyong.rhee@gmail.com}
\affiliation{Institute for Solid State Physics, University of Tokyo, Kashiwa, Chiba 277-8581, Japan}
  
  \author{S. Haas}
\email[]{shaas@usc.edu}
\affiliation{Department of Physics and Astronomy  
University of Southern California,
 Los Angeles, CA 90089-0484}
 \affiliation{School of Engineering and Science, Jacobs University
  Bremen, Bremen 28759, Germany} 
  
   \author{R. N. Bhatt}
\email[]{ravin@exchange.Princeton.EDU}
\affiliation{  Dept. of Electrical Engineering, Princeton University, Princeton, New Jersey 08544, USA}
  
  \author{S. Kettemann}
\email[]{s.kettemann@jacobs-university.de}
\affiliation{School of Engineering and Science, Jacobs University
  Bremen, Bremen 28759, Germany}  
\affiliation{Division of Advanced Materials Science, Pohang University
  of Science and Technology (POSTECH), Pohang 790-784, South Korea}

\date{\today}
\begin{abstract}
 We investigate the magnetic susceptibility $\chi(T)$ of
 quantum spin chains of $N=1280$ spins with power-law  long-range antiferromagnetic couplings
  as a function of their spatial decay exponent $\alpha$ and cutoff length $\xi$. The calculations are based on the strong disorder renormalization method which is used to obtain the temperature dependence 
  of $\chi(T)$ and distribution functions of couplings at each renormalization step. 
   For the case with only algebraic decay ($ \xi = \infty$) we find a crossover at $\alpha^*=1.066$ between a phase with a divergent low-temperature susceptibility 
   $\chi(T\rightarrow 0) $ for $\alpha > \alpha^*$
    to a phase with a vanishing $\chi(T\rightarrow 0) $ for $\alpha < \alpha^*$.
       For finite cutoff lengths $\xi$, 
      this crossover occurs at a smaller $\alpha^*(\xi)$. 
       Additionally we study the localization of spin excitations for  $ \xi = \infty$ by evaluating 
       the distribution function of excitation energies and we find a delocalization 
        transition that coincides with the opening of the pseudo-gap at $\alpha_c=\alpha^*$.

\end{abstract}

\maketitle

The  magnetic susceptibility
 of  doped semiconductors such as P-doped Si is 
 known to diverge at 
   low temperature with an anomalous power law\cite{loehneysen}.
 This can be taken as evidence for 
    local magnetic   moments,  formed  in localized states due to  interactions \cite{pwanderson0,pwanderson,mott,finkelsteinesr}, that are positioned randomly, and  
     coupled by exchange interactions \cite{andres,bhattlee,bhattreview,milovanovic},
      as illustrated in Fig. 1. 
      At low dopant density   $n_{\rm D}$       these magnetic moments 
           are coupled weakly by  the antiferromagnetic exchange 
           interaction $J$ between the hydrogen-like dopant levels \cite{andres,bps}. 
  For $n_{\rm D} \ll 1/a_B^3,$ where $a_B$ is  the Bohr radius of the dopants,
 the magnetic susceptibility    is observed  to follow 
      the Curie law $\chi \sim 1/T$ of free magnetic moments\cite{andres,sarachik,lakner}. 
     However, as the density of dopants is increased, 
 the magnetic susceptibility diverges like 
     $\chi \sim T^{-\alpha_m}$ with  a decreasing   anomalous power $\alpha_m<1.$
 This has been indentified as being  a consequence of    a random distribution of 
      exchange couplings due to   the random positions of  dopants \cite{bhattrice,rosso}. 
Indeed, in  Ref.   \onlinecite{bhattlee} it was argued that such
 random  antiferromagnetically coupled $S=1/2$ spins  
 form a ground state of 
   hierarchically coupled singlets, the random singlet phase. 
 The  random distribution of 
  excitation energies   leads to a temperature dependent
   concentration of free magnetic moments  $n_{\rm FM}(T),$ 
    resulting in an anomalous power $\alpha_m<1$.  
    
  \begin{figure}[t]
\centering
\includegraphics*[width=8cm,angle=0]{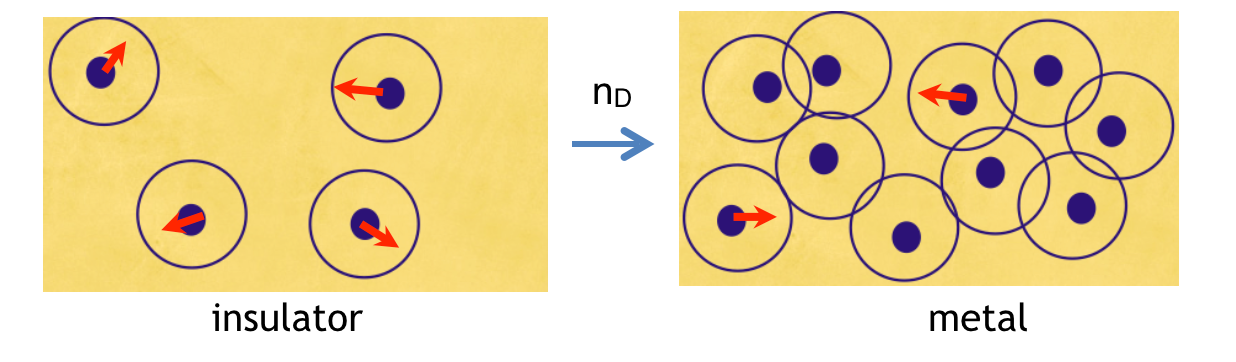}
\caption{Left: Sketch  of  electronic orbitals 
 at low doping concentration $n_{\rm D}$. All states are localized  and  magnetic,  as indicated by red arrows. Right: At larger  $n_{\rm D}$  
 states at   Fermi energy are delocalized, coexisting with 
   localized magnetic states.   }
\label{suscsip}
\end{figure}

   With increasing  doping concentration     the  density of magnetic moments 
     is observed to decrease.
  Surprisingly, both magnetic susceptibility and specific heat measurements indicate that 
  there remains a finite density of magnetic moments 
    at the metal-insulator transition \cite{loehneysen}.
    In P:Si  
    about 10\% of all dopants are magnetic at the metal-insulator transition (MIT) \cite{loehneysen}.
     These moments can be created  from localized states  in the tail of the band 
      with onsite interactions 
     \cite{mott,milovanovic}, or they may be created  due to an instability of the 
      disordered Fermi liquid with long-range interactions
     \cite{finkelsteinesr}. 
    At the 
    same time, the power $\alpha_m$ of the divergence of the magnetic susceptibility
   is experimentally observed
      to converge to a constant value as the doping concentration is increased beyond the MIT
       (in Si:P  $\alpha_m \rightarrow .64$ \cite{sarachik},  $\alpha_m \rightarrow .62$ \cite{bps}
       and $\alpha_m \rightarrow .5$ \cite{lakner,loehneysen}).
 This situation has been  modeled by a phenomenological two-fluid model \cite{andres,bhattlee,paalanen,sachdev}.
 It
           presumes that 
           the  antiferromagnetic interaction between localized magnetic moments
            is dominant even in the metallic phase, leading to 
              the formation of a 
          random singlet phase. 
            
On the metallic side of the transition 
       the  indirect exchange  interaction  $J_{ij}$ becomes long-ranged, 
         mediated by the itinerant electrons \cite{RKKY}.
             The 
            typical value of  this RKKY coupling $J_{ij}$ decays with a 
             power law with exponent $\alpha=d$, oscillating in sign 
             with a period equal to  the Fermi wavelength $\lambda_F$.
   Its amplitude  is  widely, log-normally distributed \cite{lerner}.

 Thus, aiming to get a better understanding of the magnetic properties of 
  doped semiconductors, we 
  consider the Hamiltonian 
  of  $N$ long range interacting $S=1/2$ spins,
  \begin{equation}\label{H}
H=\sum_{i\neq j, \beta}J_{ij }^\beta S_i^\beta\,S_j^\beta, 
\end{equation}
randomly  placed on a periodic lattice of length $L$
     and lattice spacing $a$. 
We 
     assume that  the   couplings   between all pairs of sites $i,j,$  are antiferromagnetic      
           decaying as 
            \begin{equation} \label{jcutoff}
J_{ij}^\beta = J|({\bf r}_i-{\bf r}_j)/a|^{-\alpha} \exp (-|{\bf r}_i-{\bf r}_j|/\xi ),
\end{equation}
which is cut off exponentially by the  length scale $\xi$,
        allowing us  to  tune between the
           limit of short-ranged coupled spins for small $ \xi  \rightarrow L/N^{1/d},$
           and the  long-range RKKY-type coupling as $ \xi \rightarrow \infty$ \cite{footnote1}.

 {\it Random Spin Chains.--}
 
 While the situation in three-dimensional systems for experimentally accessible temperatures has been at least semi-quantitatively explained, the generation of ferromagnetic bonds at intermediate stages in three-dimensions \cite{bhattlee} has made understanding the asymptotic low-energy (low-temperature) behavior not feasible. 
 
 Therefore, we focus here on the study 
 of random spin chains with long range interactions modeled by  Eq.  (\ref{H}).
 From a numerical perspective, the one-dimensional models offer the possibility of exploring larger length scales, and hopefully clearer asymptotic behavior. 
 It is well known that in one-dimensional models with power-law hopping \cite{fyodorov1996,zhou}, as well as power-law correlated disorder \cite{moura,izraelev},
  an Anderson localization-delocalization transition is found for non-interacting electrons 
  which has critical properties similar to the ones observed in the 
  three dimensional Anderson model \cite{leeramak,eversmirlin}.  Thus, one expects also a delocalization transition in random spin chains with long range interactions, as recently observed in Ref. \onlinecite{ours}.
  Moreover, 
  it is well known that many aspects of higher dimensional short ranged models can be captured in one-dimension by considering longer ranged interactions. Consequently, \textit{e.g.}, in classical spin models of ferromagnets as well as spin glasses with random ferromagnetic and antiferromagnetic bonds \cite{KotliarAndersonStein,bhattyoung}, there is a clear correspondence between the power law exponent ($\alpha$) of the one-dimensional model with the dimensionality of the higher dimensional system regarding their critical behavior.
   
\begin{figure}[t]
\begin{center}
\includegraphics[width=\columnwidth]{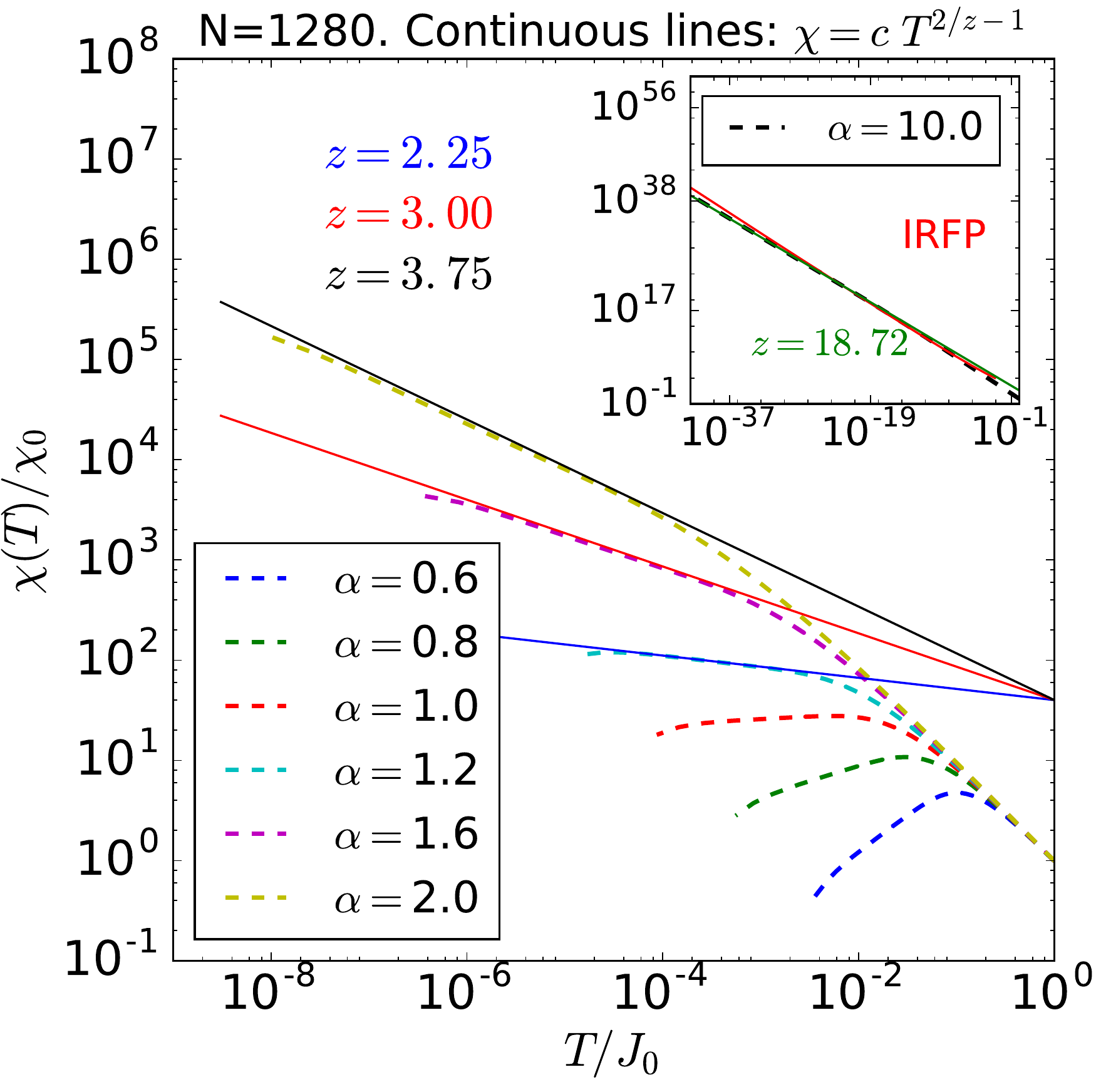}
\vspace*{-1cm}
\end{center}
\caption{Magnetic susceptibility of a chain with $N=1280$ randomly placed spins, interacting via antiferromagnetic long-range couplings $J_{ij}^x$, given by Eq. (2), with chain length  $L/a=100N$, and $\alpha=0.6, \dots, 2.0$. 
 The cutoff length is set to $\xi=\infty$.The susceptibility is normalized to its value at $T=J_0$. The continuous lines represent fits to $\chi\sim T^{2/z-1}$ with finite $z$ at low temperatures and support our argument that our model does not flow to the IRFP. The inset shows the susceptibility for $\alpha=10.0$ along with the power law with     $z=18.72$ (green line) and the IRFP result (red line).}
\label{fig:LatticeSusc}
\end{figure}   
   
The magnetic susceptibility at low temperatures is
  determined by the concentration of free paramagnetic  moments 
   $n_{FM}(T)$ (we  set $k_B=1$) \cite{bhattlee}, 
\begin{equation} \label{chi}
\chi(T)\propto \frac{n_{FM}(T)}{T}=\frac{n_M}{T}\,\int_0^Td\epsilon\,\rho(\epsilon),
\end{equation} 
where 
$n_M$ is the total density of magnetic 
 moments in the chain,  and $\rho(\epsilon)$  the density of states of spin excitations with energy $\epsilon$.  
 
 In order to compute $n_{FM}(T)$ numerically, we apply
    the  strong-disorder renormalization group (SDRG) 
             procedure \cite{bhattlee,dasguptama,fisher2}, choosing
           the pair with   largest coupling $(l,m)$
            which in its ground state 
         forms a singlet.        
               Taking the expectation value of the Hamiltonian 
        in that singlet state and performing second-order perturbation 
             theory in the coupling between 
              all spins and the spins of that singlet pair \cite{bhattlee,dasguptama,fisher,fisher2,sigrist,monthus}, we obtain
             renormalized couplings 
              between spins $(i,j)$ \cite{ours},
              \begin{eqnarray} \label{jeff}
               (J^{x}_{ij})' &=&   J_{ij}^x - \frac{(J^x_{il}-J^x_{im})(J^x_{lj}-J^x_{mj})}{J^x_{lm}+J^z_{lm}},  \nonumber \\ 
               (J_{ij}^{z})' &=&   J^z_{ij} - \frac{(J^z_{il}-J^z_{im})(J^z_{lj}-J^z_{mj})}{2 J^x_{lm}}.
              \end{eqnarray}
        
  We  implement   the  SDRG \cite{monthus}
  by  iterating these RG rules   for each realization of  bare coupling parameters
   until the system has  reached the energy $ \Omega =T$.
    We then record the number of remaining spins which have not yet formed
     a singlet, obtaining the  density $n_{FM}(T)$ \cite{bhattlee}.
    We resort to numerical iteration
   with a   large number  ($\sim$20 000) of  random realizations  needed for reliable statistics.

\begin{figure}[t]
\begin{center}
\includegraphics[width=\columnwidth]{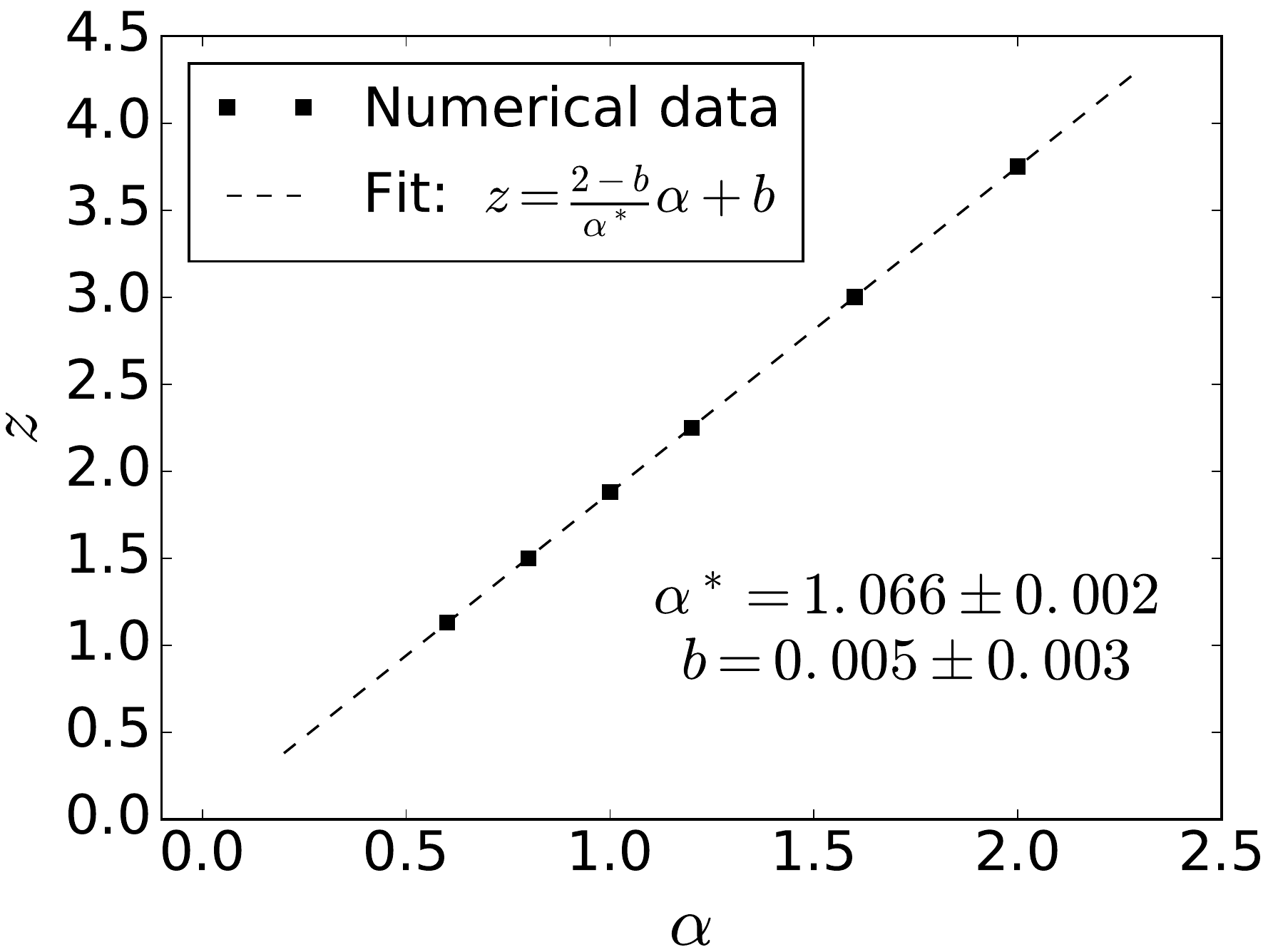}
\vspace*{-1cm}
\end{center}
\caption{Dynamical exponent $z$ extracted by fitting the low-temperature susceptibility in Fig. \ref{fig:LatticeSusc} to Eq. (\ref{susc_powerlaw}) as a function of the power $\alpha$ (squares). These numerical results are then fit to Eq. (\ref{z(alpha)}) (dashed line), which allows us to extract the crossover value $\alpha^*=1.066\pm0.002$. }
\label{fig:zVSalpha}
\end{figure}

In Fig. \ref{fig:LatticeSusc} we show
   numerical results for the susceptibility of
   the  long-ranged, $\xi = \infty$, XX-spin chain. Note that   
    the lowest temperature scale that can be 
     reached 
     for the finite system size $L$ is of the order of 
        $T_{min} =  J_{min}/k_{\rm B} = J_0 (L/2a)^{-\alpha}$, which is why the data for different values of $\alpha$  terminate at different values of $T/J_0$. 
  At low temperatures, we can see a power law behavior, which appears linear on a double logarithmic scale,  consistent with a finite dynamical exponent $z$. 
  We note that in each RG step, a fraction  $dn_{FM}/n_{FM} (\Omega)$ of the remaining spins at renormalization energy $\Omega=\text{max}(J)$ are taken away. Since this is due to the formation of a singlet with coupling $J=\Omega$, this fraction should equal $2P(J=\Omega,\Omega)d\Omega$, leading to the differential equation 
\begin{equation}\label{dn/domega}
\frac{d n_{FM}}{d \Omega} 
  = 2\,P( J=\Omega, \Omega)\,n_{FM} (\Omega),
\end{equation}  
where $P(J,\Omega)$ is the probability distribution of couplings $J$ at a given renormalization energy $\Omega$ \cite{fisher2}. At the IRFP this distribution is known to be given by
 \begin{equation}\label{P_J}
 P(J,\Omega)=(J/\Omega)^{1/z-1}/(z\Omega),
\end{equation}    
with the dynamical exponent $z=\ln(\Omega_0/\Omega)$ for initial renormalization energy $\Omega_0$. Then,  the  solution of Eq. (\ref{dn/domega}) is $n_{FM} ( \Omega ) = 1/ \ln^2(\Omega_0/\Omega),$ which yields the IRFP magnetic susceptibility $\chi(T)\sim 1/(T \ln^2(T))$ via Eq. (\ref{chi}) \cite{fisher2}. However, if the dynamical exponent $z$ is finite and fixed, the solution of Eq. (\ref{dn/domega}) in conjunction with Eq. (\ref{chi}), gives rise to a power law behavior for the low temperature susceptibility of the form 
\begin{equation}\label{susc_powerlaw}
\chi(T)\sim T^{2/z-1},
\end{equation}
consistent with our numerical results shown in Fig. \ref{fig:LatticeSusc} for $z=z(\alpha)$, a monotonically increasing function of $\alpha$ that can be extracted by linear regression fits of the susceptibility in a logarithmic scale (continuous lines).  If $z>2$, the magnetic susceptibility diverges as $T \rightarrow 0,$ with an anomalous power $\alpha_m=1-2/z<1$ that also grows with $\alpha$. In the region $z<2$, this power becomes negative and we have a vanishing susceptibility at zero temperature, consistent with the formation of a pseudo-gap in the density of sates. A similar behavior has been observed previously in Refs. \onlinecite{bhattlee} and \onlinecite{bhattdiscussions}. The crossover value $z=2$, where the susceptibility saturates to a constant, occurs at a given $\alpha=\alpha^*$, which from Fig. \ref{fig:LatticeSusc} can be concluded to be somewhere between $1.0$ and $1.2$. Assuming a linear dependence on $\alpha$ of the form 
\begin{equation}\label{z(alpha)}
z=\frac{2-b}{\alpha*}\alpha+b,
\end{equation}
we find this crossover value to be $\alpha^*=1.066\pm0.002$ by a  linear regression fit of $z(\alpha)$ as shown in Fig. \ref{fig:zVSalpha} (dashed black line), where
 the error only includes the fitting uncertainty. All values of $z$ used for the fit are found by fitting the low temperature susceptibility curves to Eq. (\ref{susc_powerlaw}) as it is done in Fig. \ref{fig:LatticeSusc} for $\alpha=1.2, 1.6$, and $2.0$.\par 
The inset in Fig. \ref{fig:LatticeSusc} displays the susceptibility for $\alpha=10.0$, along with the curve given by Eq. (\ref{susc_powerlaw}) with the value $z=18.72$ predicted by Eq. (\ref{z(alpha)}), together with  the IRFP magnetic susceptibility.   We can see  clearly
 a better  agreement of the numerical results with the finite $z$ curve,
  indicating the flow to a finite $z$ fixed point
   and not to  the IRFP, as it occurs for nearest neighbor interactions. 
   
   We note that at very large $\alpha \gg 10$ we find a finite number of free moments 
  even  at the smallest renormalization energies which are accessible in the finite spin chain. In our model, spins are randomly placed in a very diluted lattice, a situation in which nearest neighbor distances bigger than one lattice spacing $a$ is highly probable. Therefore, at very large values of $\alpha$, given the power law nature of the coupling stregths, one starts with an initial distribution $P(J)$ heavily wighted near $J=0$, which might explain the above mentioned residual free moments. A thorough exploration of this important $\alpha \gg 10$ limit is left for  future studies.



\begin{figure}[t]
\begin{center}
\includegraphics[width=\columnwidth]{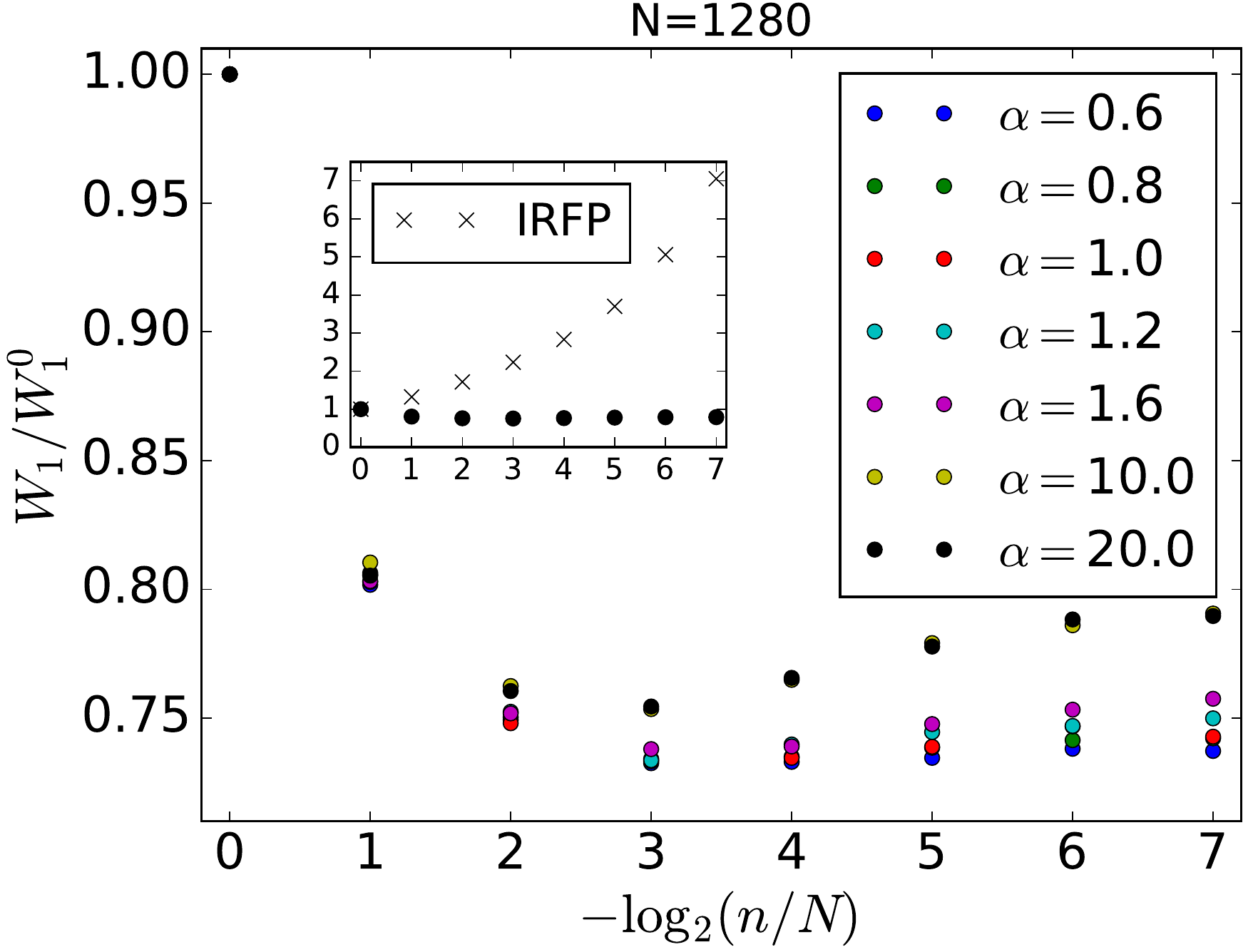}
\vspace*{-1cm}
\end{center}
\caption{Width $W_1$ of the distribution function of nearest-neighbor couplings as a function of
 the fraction of remaining spins. The negative logarithmic in base two is used to have an equally spaced horizontal variable that grows as the number of spins decreases. All values have been normalized by the width $W_{1}^0(\alpha)$ of the initial distribution and the parameters are  kept as in Fig. \ref{fig:LatticeSusc}. Inset: $W_1/W_{1}^0$ for a simple nearest neighbor model with uniformly distributed couplings (crosses) know to flow to the IRFP. The numerical results for $\alpha=20.0$ (black circles) are included for comparison purposes.   
}
\label{fig:RG}
\end{figure}

Another way to investigate whether or not there is
at finite $ \alpha$ a strong disorder fixed point with a finite dynamical exponent 
 $z(\alpha)$ or 
 a transition to 
 the IRFP at a specific finite power $\alpha_{IR}$  is to numerically inspect the evolution of the width of the couplings probability distribution with the RG flow. At the IRFP, this distribution, according to Eq. (\ref{P_J}), gets wider at every RG step, \textit{i.e.}, $W=(\langle \ln (J/\Omega_0)^2 \rangle -  \langle \ln (J/\Omega_0) \rangle^2)^{1/2}=z(\Omega) =\ln(\Omega_0/\Omega)$ increasing monotonically as  $\Omega$ is lowered 
  during the RG flow. However, as shown in Fig. \ref{fig:RG}, our system does not follow this trend for $\alpha\gg 1$ (see inset). Instead,  the width is found to   saturate to a constant value after a non-monotonic transient behavior, which is a strong indication of a finite $z$ fixed point.
  It is worth noting, that given the large number of couplings present in our system ($N(N-1)/2$ before any renormalization is performed), we have only picked the largest coupling to every spin $J_1$ in order to calculate $P(J),$ denoting the width of this approximate distribution by $W_1$.

In a previous study of the $\xi = \infty$ limit,
 we  found evidence for  a delocalization transition of spin excitations
   at a critical
 power  $\alpha_c$  by examining 
   the distribution function of the 
   lowest  excitation energy from the  ground state of long-range coupled
    random spin chains (N=128) \cite{ours}.  At $\alpha = \alpha_c$, this  gap   distribution 
    was observed to coincide with a critical function,
   separating a phase with localized  excitations at large  $\alpha > \alpha_c$,   
   where the distribution  is   Poissonian,
       from a phase with extended  
    excitations at small $\alpha < \alpha_c$, where the gap distribution follows the Wigner surmise \cite{footnote2}. 
     Since in our present study we find strong evidence in $\chi(T)$  that the density of states of spin excitations 
      presents
       a pseudogap for $\alpha < \alpha^*$,
        we revisit the gap distribution function to check if the delocalization of spin excitations at
          $\alpha_c$ coincides with $\alpha^*$.
    Following the procedure   
     carried out in Ref. \onlinecite{ours},  we now place spins randomly 
       on  the sites of  a lattice with  lattice constant $a$,
       as done  in  the calculation of the susceptibility above, and study the distribution of excitation energies. 
     
     Before proceeding, it is worthwhile to recall the results of 
     Refs. \onlinecite{fisheryoung,juhasz}, where the distribution of 
                  excitation gaps $\epsilon_1$
                   from the ground state was derived
                   for the random transverse Ising model.
                  Since  the probability to find a gap $\epsilon_1$ 
                     is proportional to the number of remaining 
                      spins $N_{FM} = n_{FM}(\Omega ) L$, at 
                      RG energy $\Omega $,
                       the distribution function of  the lowest   excitation energy $\epsilon_1$
                        equal to the 
                        energy scale of the last RG step was derived  by a scaling argument. 
                               Using the same 
                       argument for our model we obtain 
                        that the distribution of the excitation energies $\epsilon_1$
                         should have the form of a Weibull function \cite{juhasz,weibull}, 
                         \begin{equation} \label{weibull}
                         P_W(\epsilon_1) = \frac{2 u_0^{2/z} L }{z} \epsilon_1^{2/z-1} \exp (- (u_0 \epsilon_1)^{2/z} L ),
         \end{equation}
         where $u_0$ is a constant. The average excitation energy scales with 
          system size $L$ as $\langle \epsilon_1 \rangle = \frac{\Gamma(1+z/2)}{u_0} L^{-z/2}.$
 Since delocalization causes level repulsion,  Eq. (\ref{weibull})
 yields a delocalization transition when $ z (\alpha)  < z_c = 2.$         
      Thus, if this  scaling scheme of the strong disorder
       RG holds at the delocalization transitions, we conclude
        that $ z_c = 2 = z^*$ which means that the first appearance of a pseudogap coincides with the delocalization transition.  

\begin{figure}[b]
\begin{center}
\includegraphics[width=\columnwidth]{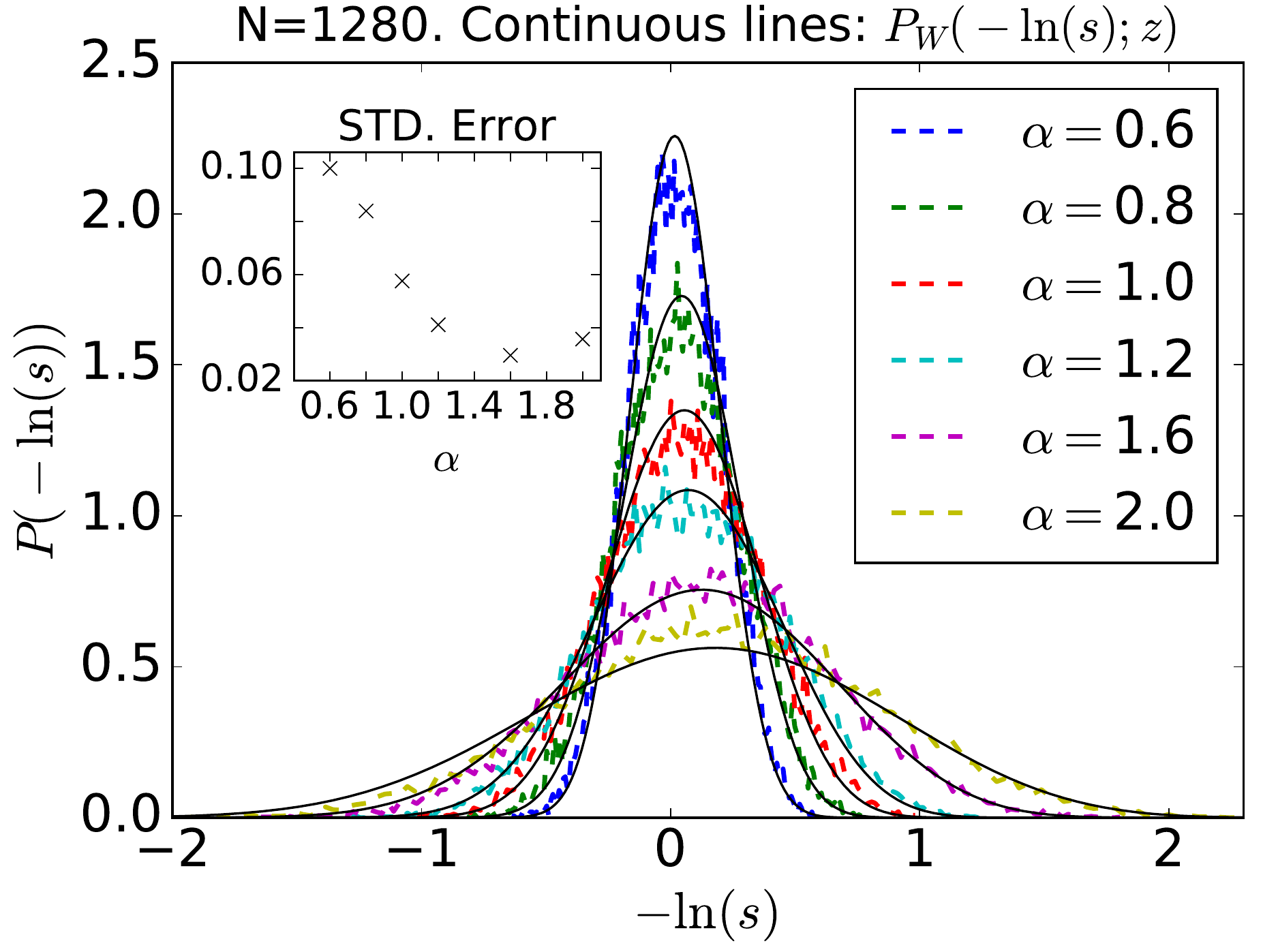}
\vspace*{-1cm}
\end{center}
\caption{Distribution of the lowest excitation gap $ \epsilon_1$  scaled by its mean value,  $ s= \epsilon_1/\langle \epsilon_1 \rangle$, for  $\xi=\infty$ and
   $\alpha=0.6, 0.8, \dots, 2.0$.
   The remaining parameters are  as in Fig. \ref{fig:LatticeSusc}. }
\label{fig:LatticeDist}
\end{figure} 

            In Fig. \ref{fig:LatticeDist}
             we show the  distribution function of the lowest excitation 
               energy $\epsilon_1$  using the logarithmic variable $x=-\ln\left(\epsilon_1/\langle \epsilon_1 \rangle\right)$ in the limit of  long-range interaction,  \textit{i.e.},   $\xi = \infty$. The continuous black curves correspond to the fits to the  Weibull distribution in Eq. (\ref{weibull}) multiplied by the  cutoff function introduced in Ref. \onlinecite{ours}, $\exp(c/(x-x_{max}))$,  which   is introduced in order to account for the fact that at finite size $L$ with periodic boundary conditions we have a maximum value $x_{max}=-\ln(\epsilon_{min}/\Delta(\alpha))$ arising from the minimal energy scale  $\epsilon_{min}=(1/2)\,J_0\,(L/6a)^{-\alpha}$. Here, the factor 6 is included since the numerical data is obtained in the third to last RG step. That way  we tried to minimize the effect of that sharp cutoff due to the finite size of the system. We found $c=16$ to work for all fits independent of the value of $\alpha$, while $u_0$ was freely changed for each curve. Given the good quality of the fits and the fact that we used the values of $z(\alpha)$ obtained from the  fitting of the susceptibility data as  plotted in Fig. \ref{fig:zVSalpha}, we can conclude that indeed, the delocalization transition occurs at the same value at which the pseudogap appears, \textit{i.e.}, $\alpha_c=\alpha^*=1.066\pm 0.002$.

       \begin{figure}[b]
\begin{center}
\includegraphics[width=\columnwidth]{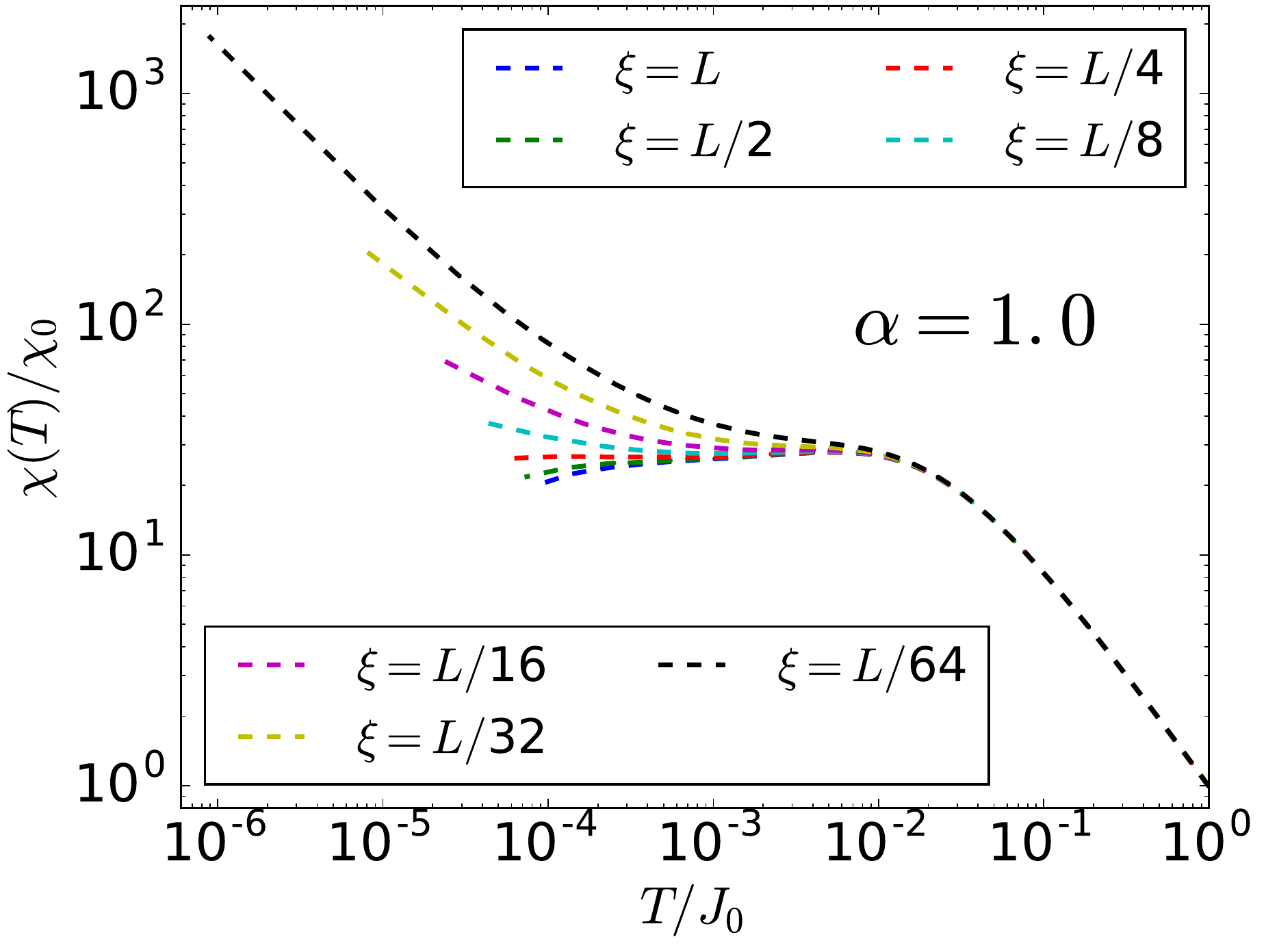}
\vspace*{-1cm}
\end{center}
\caption{Log-Log plot of the magnetic susceptibility for $\alpha=d=1$ and $\xi/L=1,1/2,\dots,1/64$. The remaining parameters are kept as in Fig. \ref{fig:LatticeSusc},
and  the susceptibility is rescaled in the same manner.}
\label{fig:YukawaSusceptibilty}
\end{figure}

As the power $\alpha=d=1$ corresponds to the typical decay of the RKKY 
 coupling in a 1D electron system
  in  the metallic regime, we may conclude from the results  in Fig. 2
   that the magnetic susceptibility 
  due to the randomly coupled magnetic moments decays to zero in the metallic regime.
 Turning on a finite cutoff $\xi,$  as caused by the finite electron localization length when the magnetic moments are
 surrounded by an electronic system,
   we see in Fig.  \ref{fig:YukawaSusceptibilty} that 
 the magnetic susceptibility 
  diverges  for $\xi<L/4.$ At low temperatures we observe a power law behavior 
  indicating a finite $z$ fixed point.  
 Increasing the cutoff to   $\xi>L/4$ we observe  in Fig.  \ref{fig:YukawaSusceptibilty}
  a  low-temperature suppression of the magnetic susceptibility,
 clearly demonstrating  the  opening of a pseudogap as the 
    range  of the interaction increases.

\begin{figure}[t]
\begin{center}
\includegraphics[width=\columnwidth]{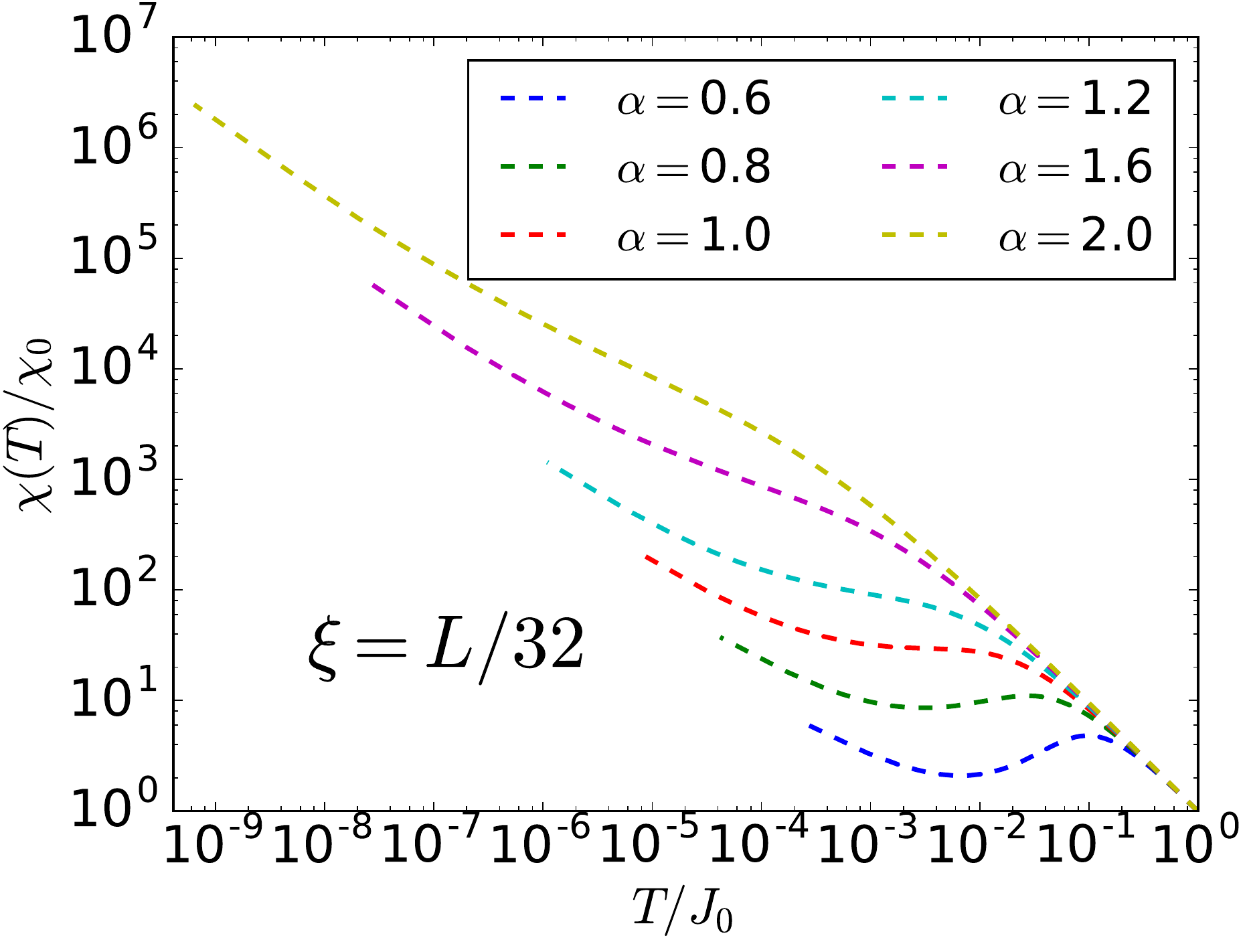}
\vspace*{-1cm}
\end{center}
\caption{Logarithmic plot of the susceptibility for a fixed and finite cutoff length $\xi=L/32$ and $\alpha=0.6, 0.8, \dots, 2.0$. The remaining parameters are kept as in Fig. \ref{fig:LatticeSusc}.}
\label{fig:FixedXiVariableAlphaSusc}
\end{figure}

For fixed small $\xi=L/32$ we observe 
in Fig. \ref{fig:FixedXiVariableAlphaSusc} 
 that at sufficiently low temperatures the magnetic susceptibility
  recovers the  power law divergence consistent with finite $z>2$, after some transient behavior. As expected, due to the presence of a finite $\xi$, this divergence is faster for every $\alpha$ when compared to the pure power-law couplings model, \textit{i.e.}, $z(\alpha,\xi=L/32)>z(\alpha,\xi=\infty)$. In fact, for $\alpha=0.6, 0.8,$ and $1.0$, we still observe
  for finite $\xi$ a divergence in $\chi(T)$, in contrast with the results shown in Fig. \ref{fig:LatticeSusc} for $\xi = \infty$, where for these values of $\alpha$ we find
   $z(\alpha)<2$ corresponding to a pseudogap. 

 In conclusion, 
 we derived the temperature dependence of the magnetic susceptibility of
 quantum spin chains with power-law  long-range antiferromagnetic couplings
  as function of the exponent $\alpha$ and the cutoff length $\xi$.
   We identified  a crossover between a phase with a divergent 
   low-temperature magnetic susceptibility
    to a phase with a vanishing low-temperature susceptibility 
     at a critical $\alpha^*$. For finite cutoff lengths $\xi$, 
      this crossover occurs at smaller values $\alpha^*(\xi) < \alpha^*$. 
       We also explored the localization of spin excitations in the limit $\xi = \infty$, by computing 
       the distribution functions of renormalized couplings and identified  a delocalization 
        transition at $\alpha_c$, which  turns out to 
         coincide  with $\alpha^*$.
         
         In order to  analyze experimental 
   results in doped bulk semiconductors the 
   study of  higher-dimensional random spin
    systems with long-range couplings is needed. However, in higher dimensions
    it  is known that even if the initial distribution 
     is purely antiferromagnetic,  ferromagnetic couplings can be 
      generated upon renormalization \cite{sigrist}. This is expected
      to modify strongly the temperature dependence of the magnetic susceptibility.
Furthermore,  as the density of itinerant electrons
 increases with the doping concentration
 the indirect exchange coupling   competes with the Kondo effect
 which screens the local moments with the itinerant electron spins. 
 Indeed, on the   metallic side of the transition  in P:Si 
 there are indications of Kondo correlations in  thermopower measurements \cite{schlager}. 
 It has been shown  that the Kondo temperature  $T_{K}$ 
 is widely distributed in the vicinity of the AMIT, 
 which  results in a power law divergence
 of the magnetic susceptibility \cite{bhatt92,langenfeld,dobros,cornaglia,kats}.
 Its power $\alpha$ has been related to 
 multifractal correlations, yielding in $d=3$ dimensions 
 with the multifractality parameter $\alpha_0$, 
 $\alpha = 2- \alpha_0/3 = .651(.652,.650)$ \cite{kats}, which 
  happens to be  close
  to the experimentally observed value \cite{lakner,bps,sarachik,loehneysen}. 
  It remains  a challenge 
  to study the  effect of the interplay of both the long range exchange couplings
   and the Kondo couplings on the 
low temperature magnetic properties.  

This research has been 
supported by DFG KE-15 Collaboration grant. H.Y. Lee acknowledges support from MEXT as Exploratory Challenge on Post-K computer (Frontiers of Basic Science: Challenging the
Limits). S. Haas acknowledges funding by 
DOE Grant Number DE-FG02-05ER46240 and 
would also like to thank the Humboldt Foundation for support. R.
N. Bhatt acknowledges support from DOE Grant No. DE-SC0002140, and
during the writing of the manuscript, the hospitality of the Aspen Center
for Physics.

Computation for the work described in this paper was supported by the University of Southern California's Center for High-Performance Computing (hpc.usc.edu).


\begin{thebibliography}
\bibitem{footnote2} We note that these  fitting functions had to be multiplied by a cutoff distribution of the form $e^{c/(x-x_{max})}$ ($x=-\log(s)$), where $ c>0$, to account for the finite size of the system giving rise to a minimum excitation energy of the order of $s_{min}=(L/2)^{-\alpha}/\Delta$ ($x_{max}=-\log(s_{min})$), where $\Delta\equiv\Delta(\alpha)$ is the average excitation gap.
\end{thebibliography}
\end{document}